\documentclass[12pt]{iopart}
\usepackage{iopams}
\usepackage{bm}

\begin{document}

\title[ Multichannel scattering problem]{Multichannel Scattering Problem  with Non-trivial Asymptotic Non-adiabatic Coupling }

\author{S L Yakovlev$^1$, E A Yarevsky$^1$, N Elander$^2$, A K  Belyaev$^3$}

\address{${^1}$ Department of Computational Physics, St Petersburg State University, 199034, St. Petersburg, Russia}
\address{${^2}$ Department of Physics, Stockholm University,  Alba Nova University Center, SE 106 91, Stockholm, Sweden}
\address{${^3}$ Department of Theoretical Physics, Herzen University,  191186, St Petersburg, Russia}

\ead{s.yakovlev@spbu.ru} 

\begin{abstract}
  The multichannel scattering problem  in an adiabatic representation is considered.
  The non-adiabatic coupling matrix is assumed to have
  a non-trivial constant asymptotic behavior at large internuclear separations. The asymptotic solutions at large internuclear distances are constructed. It is shown that these solutions  up to the first order of perturbation theory are identical to the asymptotic solutions of the re-projection approach, which was proposed earlier as a remedy for the electron translation problem in the context of the Born-Oppenheimer treatment.
\end{abstract}


\submitto{\JPA}

\maketitle

\section{Introduction}
\subsection{Background}

The adiabatic approach is one of the most widely used methods for a theoretical study of a low-energy quantum collision problem in atomic and molecular physics \cite{MM}. In the context of the two-atom collision problem, this consists of using the eigenfunctions of an electronic sub-Hamiltonian as the basis for expanding the total wave function, which then leads to the resulting multichannel equations.

One of the main approximations within this approach includes  the Born-Oppenheimer approximation \cite{BO}, which corresponds to neglecting all  non-adiabatic couplings.
The adiabatic representation is complete if all the couplings are kept in the formalism and, therefore, yields in principal
the exact solution to the Schr\"odinger equation.
The main, and computationally challenging, difficulty of solving the scattering problem in the adiabatic representation is related to the strong non-adiabatic couplings between the channels.
Critically, these may happen
in the region of an (avoided) crossing of electronic energy levels which, as a rule,
appear at finite inter-atomic distances.
One of the methods to overcome that difficulty  is by a reformulation of the problem into the diabatic representation or into the more flexible  split diabatic representation (see, for example, \cite{ThrularandMead,Esry} and references therein). Another source of non-trivial
non-adiabatic couplings is associated
with the so-called molecular-state problem (sometimes also called the electronic translation problem), which appears in the formalism as the non-zero limit of the non-adiabatic coupling matrix elements when the internuclear distance tends  to infinity
\cite{DelosandThorson1979, Delos1981, McCarroll1981, MaciasRiera1982, Belyaev1999, Belyaev2001}.
It is well understood \cite{Belyaev1999, Belyaev2001, Belyaev2002, Belyaev2010} that the physical reason for these non-vanishing asymptotic couplings is due to the fact that the adiabatic approach is based on the molecular representation and, hence, on the molecular coordinates, typically, the Jacobi molecular coordinates, in which electrons are measured from the centre of nuclear mass.
The problem is that the molecular coordinates which are used to describe fixed-nuclei molecular states of the collision complex at small and intermediate distances are not suited for the description of the free atoms in the asymptotic region.
The difference arises from the fact that the inter-atomic vectors connecting the centres of mass of colliding atoms do not coincide with the internuclear vectors which connect the centres of mass of the respective nuclei.
This results in non-zero asymptotic couplings in coupled channel equations calculated in the molecular representation.
Thus, non-zero asymptotic couplings in coupled channel equations are fundamental features of the standard Born-Oppenheimer approach.
This property, if present, makes the simple Born-Oppenheimer asymptotic form of the scattering wave function in the adiabatic representation no longer valid.

One of the
successful  methods for constructing a suitable asymptotic form of the wave function in the adiabatic representation is provided by the re-projection procedure \cite{Belyaev1999, Belyaev2001, Belyaev2002, Belyaev2010}. This procedure utilises the physically motivated asymptotic form of the total wave function which here is represented in the channel specific Jacobi coordinates by the atomic eigenstates of non-interacting atoms. The asymptotic form of the wave function in the adiabatic representation is then calculated by re-projecting this total asymptotic wave function onto the adiabatic (molecular) basis.
Although this procedure is well established,
with respect to the adiabatic representation itself it requires the use of an outer information on the asymptotic form of the total wave function.
On the other hand, a complete adiabatic representation must automatically generate the correct asymptotic form of the wave function in the adiabatic representation. The asymptotic wave function should be calculated from the asymptotic solutions of the adiabatic coupled channel equations in the region of large internuclear distances. However, as to the best of our knowledge, such a method of constructing the asymptotic wave function, which plays the role of boundary conditions in the adiabatic representation formalism, is not
properly addressed in the literature. This motivated us to undertake the present study.

The paper is organised as follows. The definition of the problem is given in the following subsection. The two channel model is considered in section two since it demonstrates the exact analytic solution of the asymptotic two-channel equations in all required details. In section three we consider the general situation when the asymptotic non-adiabatic coupling matrix couples an arbitrary number of channels $N$.
Section four concludes the paper. The proof of a technical statement about the roots of the specific polynomial with matrix coefficients appeared in the formalism is given in the appendix.

\subsection{Multichannel adiabatic equations and Born-Oppenheimer asymptotic states}\label{sub1}
The issue under consideration is the scattering problem for  the set of equations
\begin{eqnarray}
\hspace{-6mm}
\left[-\frac{d^2}{dr^2}+\frac{\ell(\ell+1)}{r^2}+{V}_j(r)-E\right]\!&F_j(r)= \nonumber\\
&\sum_{n\ge 1}\left[2{P}_{jn}(r)\frac{d}{dr}+{W}_{jn}(r)\right]\!F_n(r),
\label{eq1}
\end{eqnarray}
defined for integer $\ell\ge 0$ and real non-negative $r$, $0\le r <\infty$. The coefficients $V_j(r)$, $P_{jn}(r)$ and $W_{jn}(r)$ will be defined below.  In order to correctly formulate the problem, the set (\ref{eq1}) has to be supplied by {
appropriate boundary conditions}.
While the boundary condition at $r=0$ is natural: $F_j(0)=0$, the asymptotic ($r\to\infty$) form of $F_j(r)$ is far from trivial and depends  on the decay rate of the coefficients 
$V_j(r)$, $P_{jn}(r)$ and $W_{jn}(r)$  when $r\to\infty$.

The set of equations (\ref{eq1}) conventionally represents the Schr\"odinger equation  for a system of two atoms by using the so-called adiabatic expansion \cite{MM} for the total wave function $\Psi$ of the system. The molecular basis set $\phi_k$ generates the expansion of the total wave function
\begin{equation}
\Psi(\bm{r},\bm{\xi})=Y_\ell^{m_{\ell}}({\hat r})\sum_{n\ge 1} \frac{F_{n}(r)}{r}\phi_n(\bm{\xi},r),
\label{expansion}
\end{equation}
where $Y_\ell^{m_\ell}$ are  the standard spherical functions, for which $\ell$ and $m_\ell$ denote the total angular momentum quantum numbers, $\bm{r}$ is the internuclear relative position vector, the unit vector $\hat r$ is given by $\bm{r}/r$, where $r=|\bm{r}|$,
{while $\bm{\xi}$ represents the electronic degrees of freedom.}
The above expansion  assumes  the following form of the two-atomic Hamiltonian
\begin{equation}
H= -\frac{\hbar^2}{2M} \nabla^2_{\bm{r}}
+ h(\bm{r},\bm{\xi}),
\label{Hamiltonian}
\end{equation}
with 
$M$ denoting the nuclei reduced  mass. The sub-Hamiltonian $h(\bm{r},\bm{\xi})$ contains all interaction potentials and governs the dynamics of electrons in the field of the ``frosen" nuclei. The molecular basis set $\phi_n(\bm{\xi},r)$ is then formed by eigenfunctions of the Hamiltonian  $h(\bm{r},\bm{\xi})$
\begin{equation}
h(\bm{r},\bm{\xi}) \phi_n(\bm{\xi},r)=\lambda_n(r)\phi_n(\bm{\xi},r).
\label{phi-base}
\end{equation}
The eigenfunctions obey the orthonormality and completeness conditions
\begin{equation}\label{ortho}
\langle \phi_j|\phi_n\rangle=\delta_{jn}, \ \ \sum_{n\ge 1}|\phi_n(\bm{\xi},r)\rangle\langle \phi_n(\bm{\xi}',r|=
\delta(\bm{\xi}-\bm{\xi}'),
\end{equation}
where $\delta_{jn}$ is the Kroneker symbol and $\delta(\bm{\xi}-\bm{\xi}')$ is the delta-function.
The relative internuclear position vector $\bm{r}$ serves here as a parameter.
Throughout the paper it is
assumed that the eigenfunctions $\phi_k(\bm{\xi},r)$ and the eigenvalues $\lambda_k(r)$ depend on the magnitude $r=|\bm{r}|$ of the vector $\bm{r}$.
This property is quite common in most applications \cite{MaciasRiera1982}, and corresponds, for example, to the case when only molecular $\Sigma$ states contribute to the problem.
These assumption can also be adapted to a more general case, which involves some additional small complications to the coupling matrices
\cite{Belyaev2001}.
Finally, it is  assumed that the elements of the molecular basis $\phi_j$ are chosen to be real, which is always
possible since the sub-Hamiltonian $h(\bm{r},\bm{\xi})$ is Hermitian.

The matrices in  equations (\ref{eq1}) can now be expressed in terms of the molecular basis by the following equations:
\begin{eqnarray}
V_j(r)=\frac{2M}{\hbar^2}\lambda_j(r) \label{V}, \\
P_{jn}(r)=\frac{2M}{\hbar^2}\langle \phi_j|\frac{\partial}{\partial r}|\phi_n\rangle \label{A}, \ \
W_{jn}(r)=\frac{2M}{\hbar^2} \langle \phi_j|\frac{\partial^2}{\partial r^2}|\phi_n\rangle \label{W}.
\end{eqnarray}
Here, the brackets in matrix elements mean an integration over the electronic degrees of freedom $\bm{\xi}$.
The completeness of the basis leads to the familiar relationship  for matrices $P_{jn}(r)$ and $W_{jn}(r)$
\begin{equation}
W_{jn}(r)= \sum_{m\ge 1} P_{jm}(r)P_{mn}(r)+\frac{d P_{jn}(r)}{dr}.
\label{W-A}
\end{equation}
Since the eigenfunctions $\phi_j$ are real, the matrix $P_{jn}(r)$ is antisymmetric, i.e. $P_{jn}(r)=-P_{nj}(r)$, and, hence, it is off-diagonal $P_{jj}(r)=0$.
The constant parameter $E$ in (\ref{eq1}) represents a reduced total energy in the  colliding system.

The asymptotic behavior of the potentials $V_j(r)$ and the matrices $P_{jn}(r)$ 
is
essential for the asymptotic analysis of the solution to the equation  (\ref{eq1}).
The asymptote of $V_j(r)$ determines the asymptotic thresholds $\epsilon_j$ 
\begin{equation}\label{thresholds}
\epsilon_j = \lim_{r\to\infty} V_j(r).
\end{equation}
In this paper we assume that the potentials $V_j(r)$ are short-range, that means
\begin{equation}\label{V-sr}
\lim_{r\to\infty} r^{1+\delta}[V_j(r)-\epsilon_j]=0
\end{equation}
for  $\delta>0$. The matrix $P_{jn}(r)$ determines the asymptotic couplings $a_{jn}$ through the formula
\begin{equation}\label{couplings}
a_{jn} = \lim_{r\to\infty} P_{jn}(r).
\end{equation}
As for potentials, it is required
\begin{equation}\label{P-sr}
\lim_{r\to\infty} r^{1+\delta}[P_{jn}(r)-a_{jn}]=0
\end{equation}
for  $\delta>0$. These ``short-range" conditions lead to the following asymptotic properties of the matrices in the equations (\ref{eq1})
\begin{eqnarray}
V_j(r)\sim \epsilon_j +O(r^{-(1+\delta)}), \label{V}\\
P_{jn}(r) \sim a_{jn} +O(r^{-(1+\delta)}), \label{P}\\
W_{jn}(r) \sim \sum_{m\ge 1} a_{jm}a_{mn} + O(r^{-(1+\delta)}) \label{W}.
\end{eqnarray}
From these properties, the equations (\ref{eq1}) can be 
recast into the form
\begin{eqnarray}
\left[-\frac{d^2}{dr^2}+{\epsilon}_j-E\right]&F_j(r)= \nonumber\\
&\sum_{n\ge 1}\left[2\,{a}_{jn}\frac{d}{dr}+ \sum_{m\ge 1} a_{jm}a_{mn}  \right]F_n(r)+ Q_j(r,F)
.
\label{eq1-as}
\end{eqnarray}
In the last equation, all the terms of the order $O(r^{-(1+\delta)})$ and less have been denoted by $Q_j(r,F)$.
It follows from  formal scattering theory \cite{Newton} that terms of the order $O(r^{-(1+\delta)})$ do not affect the asymptotic
behaviour of the solution as $r\to \infty$. Therefore, in the current problem, studying the asymptote of the solution to (\ref{eq1-as}), setting $Q_j(r,F)=0$  gives the correct asymptotic form of the solution to the equation
(\ref{eq1}).

Depending on the values of the asymptotic couplings $a_{jn}$, two cases should be distinguished: ({\it i}) $a_{jn}=0$ for all $j,n$; and
({\it ii}) there exists {
 a number} $N$ such that $a_{jn}\ne 0$ for $j,n \le N$ and $a_{jn}=0$, if $j>N$ or $n>N$.
The former case ({\it i}) is somewhat ``conventional" and corresponds to the Born-Oppenheimer type of asymptotic states \cite{BelyaevPS}.
As such, due to (\ref{V-sr}) and (\ref{P-sr}), the set of equations given in (\ref{eq1-as}) becomes decoupled in the limit $r\to\infty$, and  takes the form
\begin{eqnarray}
\left[-\frac{d^2}{dr^2}+{\epsilon}_j-E\right]F_j(r)= 0.
\label{eq2}
\end{eqnarray}
The two linearly independent solutions of (\ref{eq2}) are
\begin{equation}\label{BO+-}
F^{\pm}_j(r,k_j)= \exp\{\pm i k_jr\},
\end{equation}
where the channel momenta $k_j$ are given by
\begin{equation}
\label{k}
k_j=\sqrt{E-\epsilon_j}\ge 0.
\end{equation}
These solutions  provide us with the basis for the asymptotic form of the solution to the equation (\ref{eq1}) as $r\to\infty$
\begin{equation}\label{F-as}
F_j(r) \sim k^{-{1/2}}_j[\,b^+_j F^+_j(r,k_j) + b^-_jF^-_j(r,k_j)\,].
\end{equation}
The scattering matrix can now be  defined as the transformation between incoming and outgoing amplitudes
\begin{equation}\label{S-BO}
(- 1)^{\ell+1} b^+_j=\sum_{n}S_{jn}b^-_n.
\end{equation}
The latter case ({\it ii}) is significantly more complicated, and is studied in the subsequent sections.

\section{Two channel model}\label{sec2}
As the first step in solving the problem of constructing the asymptotic form of the solution to the equation (\ref{eq1}) a model
with two channels is considered. In matrix form, this model is written as
\begin{equation}\label{2-matr-mod}
\left\{ - {\bf I}\left[\frac{{d}^2}{{d}r^2}-\frac{\ell(\ell+1)}{r^2}\right] - 2 \,{\bf P}(r)\frac{{d}}{{d}r} -{\bf W}(r)
+{\bf V}(r) - {\bf I}E
\right\}\! {\bf F}(r)=0,
\end{equation}
where ${\bf F}=\{F_1(r),F_2(r)\}^T$, the matrices ${\bf I}$, ${\bf P}$, and ${\bf V}$ are defined as
 \begin{eqnarray}
{\bf I}=&\left[\begin{array}{ccc}
1   &    0\\
0   &    1
\end{array}\right],\ \
{\bf P}(r)=
\left[ \begin{array}{cc}
0   &    p(r)\\
-p(r)   &    0
\end{array}\right],
\ \
{\bf V}=
\left[ \begin{array}{cc}
V_1(r)   &    0\\
0   &    V_2(r)
\end{array}\right]
\label{matrices}
\end{eqnarray}
and ${\bf W}(r)={\bf P}^2(r)+\frac{d}{dr}{\bf P}(r)$. {
Since the asymptotic coupling has to be non-trivial 
at $r\to\infty$, we have:}
\begin{equation}\label{p-a}
 \lim_{r\to\infty}p(r)=a\ne0.
\end{equation}
{
In contrast to (\ref{eq2}), the set of equations (\ref{2-matr-mod}) remain coupled when $r\to \infty$,  and asymptotically takes the form}
\begin{equation}
\left(-{\bf I}\frac{{d}^2}{{d}r^2} -2\,{\bf A}\frac{{d}}{{d}r} - {\bf A}^2 - {\bf K}^2\right)\!{\bf \Phi}(r)=0.
\label{SE}
\end{equation}
Here ${\bf \Phi}=\{\Phi_1(r),\Phi_2(r)\}^T$ and ${\bf A}$ and ${\bf K}^2$
are $2\times 2$ constant matrices  defined as
 \begin{eqnarray}
{\bf A}=
\left[ \begin{array}{cc}
0   &    a\\
-a   &    0
\end{array}\right],\ \
{\bf K}^2=
\left[ \begin{array}{cc}
k^2_1   &    0\\
0   &    k^2_2
\end{array}\right].
\label{I-A}
\end{eqnarray}

The next step is to construct a set of linearly independent solutions to equation (\ref{SE}).
These solutions we represent  by the following   expression for components
\begin{equation}
\Phi_j(r)=c_j\exp\{iq r\}, \ \ j=1,2,
\label{phi_k}
\end{equation}
where $q$ and $c_j$ should be determined such that the vector-function ${\bf \Phi}(r)$ {
obeys equation (\ref{SE}).}
With $\phi_j$ expressed in this form, the set of differential equations (\ref{SE}) is reduced to the following set of linear algebraic equations
\begin{eqnarray}\label{c-eq}
(q^2-k_1^2+a^2) c_1 + i2 a q c_2=0, \nonumber \\
(q^2-k_2^2+a^2) c_2 -i2 aq c_1=0.
\end{eqnarray}
The parameter $q$ should now be defined in such a way that this set of homogeneous equations has a non-trivial solution. This can be achived if the determinant of the system (\ref{c-eq}) is zero,  leading to the following  bi-quadratic equation
\begin{equation}
q^4-(k_1^2+k_2^2+2a^2)q^2+(k_1^2-a^2) (k_2^2-a^2)=0.
\label{bi-q}
\end{equation}
The four solutions to (\ref{bi-q}) are given by
\begin{eqnarray}
\hspace{-6mm}q_{1}^{\pm}
= 
\pm\frac{1}{\sqrt{2}}\sqrt{k_1^2+k_2^2+2a^2 + \sqrt{(k_1^2+k_2^2+2a^2)^2-4(k_1^2-a^2)(k_2^2-a^2)}},
\label{q1}
 \\
 \hspace{-6mm}q_{2}^{\pm}
 = 
 \pm \frac{1}{\sqrt{2}}\sqrt{k_1^2+k_2^2+2a^2 - \sqrt{(k_1^2+k_2^2+2a^2)^2-4(k_1^2-a^2)(k_2^2-a^2)}}.
\label{q-12}
\end{eqnarray}
The case of small   
$a$ is
of interest, since in the atomic collision problem $a$ is proportional to the
square root of the ratio of the electron to the nuclear mass.
In this case, the asymptotes of $q_n$, when $a\to 0$, can easily be evaluated
from (\ref{q1}), (\ref{q-12}) and have the form
\begin{eqnarray}
q_n^\pm
= \pm k_n \pm \frac{1 \pm 2\frac{k_1^2+k_2^2}{k_1^2-k_2^2}}
{2k_n}a^2 +O(a^4),\ \ n=1,2 .
\label{q-lim}
\end{eqnarray}
These formulae show that in the limit $a\to 0$ the quantities $q_n$ approach the momenta $k_n$.
It is important to emphasize that  $q^\pm_n-(\pm k_n)=O(a^2)$.

The set of equations (\ref{c-eq}) for two-component vectors ${\bf C}^{n\pm}$ 
 can now be solved by inserting values of $q^\pm_{n}$ into (\ref{c-eq}).
The resulting solution is represented by the following expressions:
\begin{eqnarray}
{\bf C}^{1\pm}= \left[\begin{array}{c} c^{1\pm}_{1}\\ c^{1\pm}_{2}\end{array}\right]=
N(a,q^\pm_1,k_1)\left[\begin{array}{c} 1\\ -\frac{(q^\pm_1)^2-k^2_1+a^2}{i2aq^{\pm}_1}\end{array}\right], \\
{\bf C}^{2\pm}= \left[\begin{array}{c} c^{2\pm}_{1}\\ c^{2\pm}_{2}\end{array}\right]=
N(a,q^\pm_2,k_2)\left[\begin{array}{c} \frac{(q^\pm_2)^2-k^2_2+a^2}{i2aq^{\pm}_2}\\ 1 \end{array}\right],
\end{eqnarray}
where the normalisation factor $N$ is calculated from
\begin{equation}
N(a,q,k)=
\left\{
{1+ \frac{(q^2-k^2+a^2)^2}{4a^2q^2}}\right\}^{-1/2}
%
\label{N}
\end{equation}
using the relevant  values of $q$ and $k$. As earlier for $q^\pm_n$ quantities, consider the situation 
when $a\to 0$. In this case,
the following behaviour of the components of the vectors  ${\bf C}^{n\pm}$, $n=1,2$ 
is obtained
\begin{eqnarray}
c^{n\pm}_j= \delta_{jn} \pm ia \frac{2k_n}{k_1^2-k_2^2}[1-\delta_{jn}] + O(a^2), \ \ j=1,2.
\label{c-nj-as}
\end{eqnarray}

The four solutions derived for equation (\ref{SE})
\begin{equation}
{\bf \Phi}^{n\pm}(r,q^\pm_n)={\bf C}^{n\pm}\exp\{iq^{\pm}_n r\}, \ \ n=1,2
\label{Phi-pm}
\end{equation}
are the set of asymptotic states which were necessary to construct the asymptote to the solution of the two-channel equation (\ref{2-matr-mod})
\begin{equation}
{\bf F}(r)\sim  \sum_{n=1}^2 q^{-1/2}_n
[\,b^+_{n} {\bf \Phi}^{n+}(r,q^+) + b^-_{n} {\bf \Phi}^{n-}(r,q^-_n)\,], \ \ r\to\infty .
\label{as-F-2}
\end{equation}
Analogously to the definition (\ref{S-BO}) given in  the preceding section, the $2\times 2$ scattering matrix is
introduced as the transformation matrix for incoming $b^-_j$ and outgoing $b^+_j$ amplitudes
\begin{equation}\label{S-2mod}
(- 1)^{\ell+1} b^+_j=\sum_{n=1}^2S_{jn}(a)b^-_n.
\end{equation}

The asymptotic representation (\ref{as-F-2}) can be simplified if the asymptotes (\ref{q-lim}) and (\ref{c-nj-as}) for $q_n^\pm$ and ${\bf C}^{n\pm}$ as $a\to 0$  are used in
(\ref{Phi-pm}). Neglecting  terms of the order $O(a^2)$ leads to the following form of the components of the asymptotic solution
(\ref{as-F-2})
\begin{equation}
{F}_j(r)\sim  \sum_{n=1}^2 k^{-1/2}_n
[\,b^+_{n} t^{+}_{jn}F^+_n(r,k_n) 
+ b^-_{n} t^{-}_{jn}F^-_n(r,k_n) 
\,].
\label{as-F-t}
\end{equation}
Here, the $t^{\pm}_{jn}$ coefficients are defined as
\begin{equation}\label{t-pm}
t^{\pm}_{jn}= \delta_{jn} \pm ia \frac{2k_n}{k_1^2-k_2^2}[1-\delta_{jn}], \ \ j,n=1,2.
\end{equation}
It is easily verified that formulae (\ref{as-F-t}) and (\ref{t-pm}) are identical to the representation of the asymptotic solution
that is obtained by the re-projection method (see, for instance,  formulae (18) and (21) from the paper \cite{BelyaevPS}).

It is worth noting that in the limit $a\to 0$ the coefficients $t^\pm_{jn}$ take the Kroneker-delta $\delta_{jn}$ form and, consequently,
the representation (\ref{as-F-t}) turns into (\ref{F-as}). For the scattering matrix this limit yields
$ S_{jn}(a)\to S_{jn}$ with $S_{jn}$  defined by (\ref{S-BO}).

\section{General case of multichannel equations}
In this section we consider the general situation, i.e. the case ({\it ii}) described in subsection \ref{sub1}, when $N$ states remain asymptotically coupled as $r\to \infty$.
Here, the set of equations (\ref{eq1}), (\ref{eq1-as}) is asymptotically split into two pieces:
the nontrivial $N\times N $ system for components $F_j$, $j=1, \ldots, N$; and the trivial decoupled set of the form (\ref{eq2})
for components $F_j$, $j>N$.
The latter leads to the asymptotic form of the components $F_j$, $j>N$,  given in Eq.~(\ref{F-as}).
The asymptotic form of the former components for $j=1,\ldots, N$ should be constructed by solving the relevant set of equations with non-trivial asymptotic coupling. As it was argued  in subsection \ref{sub1}, this set is obtained from (\ref{eq1}) or (\ref{eq1-as}) by neglecting all terms of the order $O(r^{-1-\delta})$ and less. The resulting asymptotic $N\times N$ set of equations  for the vector-function ${\bf \Phi}(r)=\{\Phi_1(r),\ldots,\Phi_N(r)\}^T$ is,
in the matrix form, given by:
\begin{equation}
\left(-{\bf I}\frac{{d}^2}{{d}r^2} -2{\bf A}\frac{{d}}{{d}r} - {\bf A}^2 - {\bf K}^2\right){\bf \Phi}(r)=0.
\label{SEN}
\end{equation}
The $N\times N$ matrices $\bf I$, ${\bf K}^2$ and $\bf A$ are given by their matrix elements
\begin{eqnarray}
[{\bf I}]_{jn}=\delta_{jn} , \ \ [{\bf K}^2]_{jn}=\delta_{jn}k^2_n, \ \
[{\bf A}]_{jn}=a_{jn}.
\label{matrN}
\end{eqnarray}
In the same way as discussed  in  section \ref{sec2}, by using the substitution
\begin{equation}
{\bf \Phi}(r)={\bf C}\exp\{iqr\}
\label{PhiN}
\end{equation}
with the unknown constant $q$ and vector ${\bf C}=\{C_1,\ldots,C_N\}$, the set of differential equations given in (\ref{SEN})
can be reduced to a set of linear algebraic equations:
\begin{equation}
\left(q^2\,{\bf I} -2q\,i{\bf A} - {\bf A}^2 - {\bf K}^2\right){\bf C}=0.
\label{CEN}
\end{equation}
The non-trivial solution to (\ref{CEN}) exists if and only if the determinant, which is a polynomial of degree $2N$, of that system is zero
\begin{equation}
D(q)\equiv  \det\left(q^2\,{\bf I} -2q\,i{\bf A} - {\bf A}^2 - {\bf K}^2\right)=0.
\label{detC}
\end{equation}
It is shown in the Appendix that the equation (\ref{detC}) can be satisfied only for real values of $q$.  Since the matrix ${\bf A}$ is antisymmetric with respect to the matrix transposition ${\bf A}^T=-{\bf A}$, the following property
\begin{equation}
\left(q^2\,{\bf I} -2q\,i{\bf A} - {\bf A}^2 - {\bf K}^2\right)^T=\left(q^2\,{\bf I} +2q\,i{\bf A} - {\bf A}^2 - {\bf K}^2\right)
\label{sym}
\end{equation}
shows that the left hand side of (\ref{detC}) is even function of  real $q$, i.e. $D(-q)=D(q)$. Therefore,
the $2N$ roots of equation (\ref{detC}) form the symmetric real set $\pm q_1,\ldots,\pm q_N$, such that $D(\pm q_n)=0$, $n=1,\ldots,N$. For definiteness, the signs   of $q_n$ are fixed in such a way that $q_n\ge0$, $n=1,\ldots,N$.
The $2N$ vectors ${\bf C}^{n\pm}$, $n=1,\ldots,N$ can  now be calculated as the solutions to equation (\ref{CEN}) by inserting $\pm q_n$ quantities. These quantities determine $2N$ solutions to the equation (\ref{SEN}) of the form
\begin{equation}
{\bf \Phi}^{n\pm}(r,q_n)={\bf C}^{n\pm}\exp\{\pm i q_n r\}.
\label{Phi-n}
\end{equation}

All ingredients which are necessary for constructing the asymptotic ($r\to\infty$) boundary conditions for the equation (\ref{eq1}) are now available.
The components $\Phi^{n\pm}_j$ of states (\ref{Phi-n}) for $j=1,\ldots,N$ together with components $F^\pm_j$ from (\ref{BO+-}) for $j>N$
form the basis for those asymptotic boundary conditions for the components $F_j(r)$ of the solution to the equations (\ref{eq1})
\begin{eqnarray}
F_j(r)\sim  \sum_{n=1}^{N}
{q_n}^{-1/2}\left[\,b^-_n \Phi^{n-}_j(r,q_n) + b^{+}_n\Phi^{n+}_j(r,q_n)\,\right], \ \ j=1,\ldots,N, \label{F-j-1} \\
F_j(r)\sim {k_j}^{-1/2}\left[\, b^-_j F^-(r,k_j) + b^+_j F^+(r,k_j) \, \right], \ \ j> N .
\label{F-j-2}
\end{eqnarray}
These boundary conditions are our main result that completes  the definition of  the scattering problem
for the set of equations (\ref{eq1}) in the general situation.

The respective $S$-matrix is then defined as the transformation matrix between the incoming and outgoing amplitudes in scattering channels
\begin{equation}\label{S-Nmod}
(- 1)^{\ell+1} b^+_j=\sum_{n\ge 1}S_{jn}({\bf A})b^-_n.
\end{equation}

As in the preceding section, we consider the special case when the coupling matrix ${\bf A}$ is small.
Here, it is useful to introduce
the scalar coupling constant $a$ in such a way that ${\bf A}=a{\hat {\bf A}}$  in order to explicitly characterize the order of the coupling matrix.
For small coupling constant $a\to 0$, the solution to the equation (\ref{CEN}) can be obtained perturbatively. 
${\bf C}$ and $q$ are represented as
\begin{equation}
q=q_0+aq_1 + O(a^2), \ \ {\bf C}={\bf C}^{(0)}+a{\bf C}^{(1)} + O(a^2)
\label{qC}
\end{equation}
and are substituted  into (\ref{CEN}). The resulting equations for ${\bf C}^{(0)}$ and ${\bf C}^{(1)}$ are given by
\begin{eqnarray}
(q^2_0-{\bf K}^2){\bf C}^{(0)}=0 \label{C0},  \\
(q^2_0-{\bf K}^2){\bf C}^{(1)}=-2q_0(q_1-i{\hat {\bf A}}){\bf C}^{(0)}
\label{C1}.
\end{eqnarray}
The first equation (\ref{C0}) shows that $q^2_0$ and ${\bf C}^{(0)}$ must be the eigenvalue and the eigenvector of the diagonal matrix ${\bf K}^2$, respectively.
The eigenvalues set of ${\bf K}^2$ is $k^2_1,\ldots,k^2_N$ and the corresponding eigenvectors ${\bf C}^{(0)n}$, $n=1, \ldots, N$, are given by components as
$C^{(0)n}_j=\delta_{jn}$. To solve the second equation (\ref{C1}), we take ${\bf C}^{(0)}={\bf C}^{(0)n}$ and $ q_0^2=k^2_n$ which leads to $q_0=\pm k_n$. Since $\mbox{Ker}(k^2_n-{\bf K}^2)$ is not trivial, the resolution condition
\begin{equation}
\mp 2k_n \langle (q_1-i{\hat {\bf A}}){\bf C}^{(0)n}, {\bf C}^{(0)n}\rangle=0
\label{res0cond}
\end{equation}
is required to guaranty the existence of a non-trivial solution to (\ref{C1}). Here, $\langle .,.\rangle$ means the standard inner product in
$\mathbb{C}^N$.  Evaluating the inner product gives $q_1=i{\hat a}_{nn}$, hence $q_1=0$, since ${\hat a}_{jn}$ is off-diagonal.
The equation (\ref{C1}) takes now the form
\begin{equation}
(k^2_n-{\bf K}^2){\bf C}^{(1)n\pm}=\pm 2i k_n{\hat {\bf A}}{\bf C}^{(0)n}.
\label{C1n}
\end{equation}
For the components, it is given by
\begin{eqnarray}
(k^2_n-{k}^2_j){C}_j^{(1)n\pm}=\pm 2i k_n{\hat {a}}_{jn}
  \label{C1nj}
\end{eqnarray}
for $j\ne n$, while
\begin{equation}
{C}_n^{(1)n\pm}=0
\label{C-nn=0}
\end{equation}
according  to the standard convention of the perturbation theory \cite{Landau}.
From equations (\ref{C1nj}), (\ref{C-nn=0}), it is seen that
\begin{eqnarray}
{C}_j^{(1)n\pm}=\pm \frac{2i k_n}{k^2_n-{k}^2_j}{\hat {a}}_{jn}(1-\delta_{jn}).
  \label{C1njf}
\end{eqnarray}
Finally, from (\ref{qC}), the following expressions  are obtained for $q$ and for components of ${\bf C}$
\begin{eqnarray}
q=\pm k_n +O(a^2), \ \ n=1, \ldots, N \label{q-n}, \\
C_j^{n\pm}= \delta_{jn}\pm \frac{2ik_n}{k^2_n-k^2_j} a_{jn}(1-\delta_{jn}) +O(a^2), \ \ j,n=1, \ldots, N.  \label{CC}
\end{eqnarray}
Here it has been taken into account that $a{\hat a}_{jn}=a_{jn}$.
With these representations the explicit form for the leading terms of the components of the asymptotic states ${\bf \Phi}^{n\pm}$
are obtained
\begin{equation}
\Phi^{n\pm}_j(r,q_n) = t^\pm_{jn}\exp\{\pm i k_nr\} +O(a^2)
\label{Phi-t}
\end{equation}
with $t^{\pm}_{jn}$ defined as
\begin{equation}
t^{\pm}_{jn}= \delta_{jn}\pm \frac{2ik_n}{k^2_n-k^2_j} a_{jn}(1-\delta_{jn}).
\label{t-pm-N}
\end{equation}
The respective asymptotic ($r\to\infty$) boundary conditions (\ref{F-j-1}) for components $F_j(r)$ with $j=1,..., N$  now become
\begin{equation}
F_j(r)\sim  \sum_{n=1}^{N}
{k_n}^{-1/2}\left[\,b^-_n t^-_{jn} \exp\{-ik_nr\} + b^{+}_n t^+_{jn }\exp\{ik_nr\}\,\right] +O(a^2).
 \label{F-j-a}
\end{equation}
It is noted that the latter formula is identical to the formula (18) that is obtained in \cite{BelyaevPS} within the re-projection procedure.
The $t$-matrix components (\ref{t-pm-N}) are also identical to the respective $t$-matrix components from  \cite{BelyaevPS} if the definitions (\ref{V}),  (\ref{thresholds}) and (\ref{k}) are taken into account.

The derived formulae (\ref{q-n}) and (\ref{CC}) give the leading with respect to the coupling constant $a$ terms for $q$ and ${\bf C }$.
Subsequent terms of the decompositions can be obtained (if it is necessary) by implementing the standard prescription of the perturbation theory \cite{Landau}. It is also worth mentioning that the case of degenerate asymptotic scattering channels can be treated with the relevant variant of the perturbation theory, and, therefore, may require applying additional corrections in the construction of asymptotic boundary conditions, as shown, for example, in \cite{Belyaev2015}.



\section{Conclusion}
We have presented a general formalism of  constructing the asymptotic boundary conditions for solutions to the adiabatic multi-channel
scattering problem in the case when the non-adiabatic coupling matrix remains non-trivial at large internuclear distances.
These asymptotic conditions generalise
the Born-Oppenheimer form of the asymptotic boundary conditions, which commonly  used  in the case of asymptotic decoupling of equations.
The non-zero asymptotic non-adiabatic coupling matrix elements are fundamental features of the Born-Oppenheimer approach, so the construction of the asymptotic solutions to the coupled channel equations with such couplings 
is not only of general, but also of practical importance for calculations of inelastic cross sections.
The calculations of inelastic cross sections with non-zero asymptotic non-adiabatic couplings have been successfully accomplished by means of the re-projection method for a number of collisional processes, e.g., in collisions of Li + Na \cite{Belyaev2010}, He + H \cite{Belyaev2015}, Mg + H \cite{Belyaev2012}, Li$^+$ + He and Li + He$^+$ \cite{Belyaev2015mnras}.
The asymptotic solutions (\ref{t-pm-N}), (\ref{F-j-a}) derived via the general formalism of the present paper 
are identical to the asymptotic solutions used in the re-projection method \cite{Belyaev1999, Belyaev2001, Belyaev2010, BelyaevPS} within the first order of perturbation theory.  
Therefore, numerical calculations of inelastic cross sections would lead to identical results.
Thus, there is no need to repeat any numerical calculations in the present paper.

The formulation of the scattering problem for multichannel coupled equations with the conditions derived in the present paper covers two most important cases of short-range and constant long-range
asymptotic behavior of the non-adiabatic coupling matrix.
%
%
The case of as slow as $O(r^{-1})$ vanishing of the non-adiabatic coupling matrix and/or adiabatic potentials  is of interest, particularly, for the hyper-spherical adiabatic approach \cite{Esry}.
The solution of the asymptotic boundary condition problem in that case requires some
specific new technique and will be explained elsewhere.

The derived asymptotic form of the solution to the adiabatic coupled equations opens a way for
using the recently developed potential splitting approach for solving the multichannel scattering problem  in both the diabatic  and  adiabatic  forms of the multichannel equations \cite{PRA, ChPh}.

\ack
The work of SLY and EAY  is supported by the Russian Foundation for Basic Research grant No. 14-02-00326
and by St Petersburg State University within the project No. 11.38.241.2015. The work of NE is supported by a grant from the Carl Trygger
Foundation. The work of AKB is supported by the Russian Foundation for Basic Research grant No. 16-03-00149.

\section*{Appendix}
In order to show that the equation (\ref{detC}) can possess  the nontrivial solution only for real values of the parameter $q$ 
the equivalent fact that
$\mbox{Ker}( q^2\,{\bf I} -2q\,i {\bf A} -{\bf A}^2 - {\bf K}^2)$
can be non-trivial only for real values of $q$ is proven. Begin from the 
identity
\begin{equation}
q^2\,{\bf I} -2q\, i {\bf A} -{\bf A}^2 - {\bf K}^2=(q{\bf I}-i{\bf A})^2-{\bf K}^2
\label{A1}
\end{equation}
and the fact that $i{\bf A}$ is Hermitian.
Let $q=\alpha+i\beta$ be complex and ${\bf C}\in \mbox{Ker}((q{\bf I}-i{\bf A})^2-{\bf K}^2),$
then
\begin{equation}
\langle[(q{\bf I}-i{\bf A})^2-{\bf K}^2]{\bf C},{\bf C}\rangle=0,
\label{A2}
\end{equation}
where $\langle.,.\rangle$ is the standard inner product in $\mathbb{C}^N$.
Since the left hand side of this equation is represented as
\begin{equation}
\langle (\alpha-i{\bf A})^2{\bf C},
{\bf C}\rangle
- \langle {(\beta^2+\bf K}^2){\bf C},
{\bf C}\rangle +
2i\beta \langle (\alpha -i{\bf A}){\bf C},{\bf C}\rangle
\label{A3}
\end{equation}
equation (\ref{A2}) is equivalent to the following two equations
\begin{eqnarray}
\langle (\alpha-i{\bf A})^2{\bf C},
{\bf C}\rangle
- \langle {(\beta^2+\bf K}^2){\bf C},
{\bf C}\rangle=0, \ \ \label{A21}\\
\beta \langle (\alpha -i{\bf A}){\bf C},{\bf C}\rangle=0. \label{A22}
\end{eqnarray}
If $\beta \ne 0$ and ${\bf C}\ne 0$, from (\ref{A22}) it follows that
$$
(\alpha -i{\bf A}){\bf C}=0.
$$
The latter reduces the left hand side of (\ref{A21}) to $- \langle {(\beta^2+\bf K}^2){\bf C},{\bf C}\rangle$.
The matrix
$\beta^2+{\bf K}^2$ is obviously positive defined for $\beta\ne 0$, hence $- \langle {(\beta^2+\bf K}^2){\bf C},{\bf C}\rangle <0$, which contradicts equation (\ref{A21}). As a result, it is proven that for complex $q$ with $\Im\mbox{m}\, q=\beta\ne 0$,  equation
(\ref{A2}) can be satisfied only for ${\bf C}=0$ and, consequently,  $\mbox{ Ker}((q{\bf I}-i{\bf A})^2-{\bf K}^2)$ is trivial.


\end{document}